# Freezing of short-range ordered antiferromagnetic clusters in the CrFeTi$_2$O$_7$ system

Arun Kumar[1], Soumendra Nath Panja[1], Lukas Keller[2] and Sunil Nair[1]

[1]Department of Physics, Indian Institute of Science Education and Research (IISER) Pune-411008, India

[2]Laboratory for Neutron Scattering and Imaging, Paul Scherrer Institut, CH-5232 Villigen PSI, Switzerland

**Abstract:**

We report on the CrFeTi$_2$O$_7$ (CFTO) system using a combination of x-ray diffraction, dc magnetization, ac susceptibility, specific heat and neutron diffraction measurements. CFTO is seen to crystallize in a monoclinic *P2$_1$/a* symmetry. It shows a glassy freezing at $T_f \sim$ 22 K, characterized by the observation of bifurcation between ZFC and FC χ (T) curves, frequency dispersion across $T_f$ in ac susceptibility, appearance of non-zero remanent magnetization and coercivity below $T_f$, very slow relaxation of iso-thermal remanent magnetization with time and a linear temperature dependence of magnetic contribution to specific heat C$_m$ below $T_f$. The microscopic neutron diffraction analysis of CFTO not only confirms the absence of long-range antiferromagnetic ordering but also exhibits diffuse scattering due to the presence of short-range ordered antiferromagnetically correlated spin clusters.

**Introduction:**

In the recent past, pyrochlore oxides with the chemical formula $A_2B_2O_7$ (where $A$ is a trivalent rare-earth cation and $B$ is a tetravalent transition metal cation) have attracted enormous attention because of the diverse and exotic electronic and magnetic phenomena they exhibit. This includes metal-insulator transitions [1,2], Weyl semimetal state [3], Fermi-arc surface [4], topological Mott insulators [5,6], non-Fermi liquids [7,8], giant magnetoresistance [9,10], superconductivity [11], quantum spin liquid [12], classical/quantum spin ice states [13,14], unconventional spin-glass states [15,16] and spin-phonon coupling [17,18]. Most of the rare-earth pyrochlore oxides usually crystallize in an ordered cubic structure ($Fd$-$3m$ space group), consisting of two interpenetrating lattices of corner-sharing tetrahedral frameworks of $A$ and $B$ cations [19,20]. It can also be visualized as the stacking of alternating kagome and triangular sublattices along the crystallographic [111] direction of the cubic unit cell. The presence of such fascinating structures gives rise to a strong geometrical frustration on the pyrochlore lattice [20].

Previous studies on the $A_2B_2O_7$ pyrochlore oxides were mainly focused on systems with a rare-earth cation at the $A$-site and transition metal cations like Ti, Zr, Mo, and Ir at the $B$-site [19,20]. Investigations on $A_2B_2O_7$ systems with 3d transition metal cations at both sites are limited to the vanadate $A_2V_2O_7$ system [21-34]. Recent investigations on the family of transition metal vanadates $A_2V_2O_7$ with A = Mn, Co, Ni, and Cu have discovered a wide range of interesting phenomena like ferroelectricity [21–25], multiferroicity with quantum criticality [21,26], magnetodielectricity [27–29], quantum magnetization plateaus [23,24,30], multiple metamagnetic phase transitions [31], spin-flop transitions [32], field-induced ferroelectric phases [24] and nonreciprocal low energy excitations [33]. Unlike rare-earth pyrochlore oxides, the transition metal vanadate family does not crystallizes in the pyrochlore type structure but exhibits a variety of crystallographic phases depending on the size of $A$-site cations [21,29,30,34]. $Mn_2V_2O_7$ has a distorted honeycomb layered structure and stabilizes in two different structural phases α (triclinic: $P\bar{1}$ space group) and β (monoclinic: $C2/m$ space group) in low (below < 284 K) and high-temperature (above > 310 K) regions, respectively [29]. Both $Ni_2V_2O_7$ and $Co_2V_2O_7$ crystallize in the monoclinic structure in the $P2_1/c$ space group [21,30]. On the other hand, $Cu_2V_2O_7$ crystallizes in three different polymorphs, namely, *α, β,* and *γ* phases with orthorhombic (*Fdd2* space group), monoclinic (*C2/c* space group), and triclinic ($P\bar{1}$ space group) symmetries, respectively [34].

Motivated by the intriguing functional properties of the $A_2V_2O_7$ vanadate family, it is of interest to explore the other transition metal cation-based titanate $A_2Ti_2O_7$ (with A = Cr, Fe) systems for their structure and property correlations. Historically, $Cr_2Ti_2O_7$ was first reported by Hamlin et al. [35] and its structure corresponds to a monoclinic phase based on a related α-$PbO_2$ columbite structure. In a subsequent study, it was found that this compound is difficult to synthesize due to the requirement of very high calcination/sintering temperature for several days [36,37]. This synthesis process also leads to a refractory/brittle nature of the sample. On the other hand, $Fe_2Ti_2O_7$ is still unexplored and its synthesis and properties are yet to be reported. Kwestroo et al. [37] reported that replacing Cr by Fe upto 66 % in $Cr_2Ti_2O_7$ leads to a stable compound without any structural change. The replacement of Cr by Fe also causes a decrease in the synthesis temperature. Grey et al. [36] investigated the room-temperature crystal structure of $CrFeTi_2O_7$ and showed that it corresponds to a monoclinic structure in the $P2_1/a$ space group. To the best of our knowledge, no detailed investigation of the physical and magnetic properties of $CrFeTi_2O_7$ have been performed till date.

In an attempt to understand the structure and properties of transition metal cation-based titanates, we have synthesized $CrFeTi_2O_7$ and extensively characterized it using a variety of probes like X-ray diffraction (XRD), Energy dispersive X-ray spectroscopy, dc magnetization, ac susceptibility, specific heat and neutron powder diffraction. Our results show that the ground state of *CrFeTi$_2$O$_7$* corresponds to a glassy state ($T_f \sim$ 22 K) originating from the freezing of short-range ordered antiferromagnetic (AFM) spin clusters. We believe that this study will strengthen our current understanding of the *A$_2$B$_2$O$_7$* family of transition metal-based oxides.

**Experimental Details:**

Polycrystalline specimens of *CrFeTi$_2$O$_7$* (CFTO) were prepared by a conventional solid-state route using high-purity oxides Fe$_2$O$_3$, Cr$_2$O$_3$ and TiO$_2$ (Sigma Aldrich, purity $\geq$ 99.9 %). The precursor powders in the stoichiometric ratio were mixed thoroughly in an agate mortar and pestle with ethanol as a mixing medium and pressed into pellets (13 mm diameter and 1.5 mm thickness). The pellets were loaded in a furnace and heated upto 1623 K at the rate of 6 K/min, kept there for 48 hrs and then cooled to room temperature. The surfaces of the pellets were cleaned with ethanol and then crushed gently to powder for x-ray diffraction and neutron scattering measurements.

The chemical compositions of the sample were checked by the energy dispersive x-ray (EDS) spectroscopic technique (Oxford Inca Detector) attached in the scanning electron microscope (Zeiss Ultra Plus). The room temperature x-ray diffraction (XRD) measurements were performed using a high-resolution powder diffractometer (Bruker, model no. D8 ADVANCE) with CuK$\alpha$ source ($\lambda$ = 1.5406 Å). Analysis of the XRD data were performed by Rietveld refinement using Fullprof software [38]. Temperature (T), field (H), time (t) and frequency(f)-dependent magnetic measurements were carried out using a SQUID-based magnetometer (MPMS-XL, Quantum Design Inc.). Heat capacity as a function of temperature was measured by the relaxation method using a physical property measurement system (PPMS, Quantum Design Inc.) with HC module. Neutron powder diffraction (NPD) experiments were performed using the cold neutron diffractometer DMC with an incident wavelength of 2.458 Å at the Paul Scherrer Institute (PSI) Facility, Switzerland [39,40]. The powder sample was loaded in a cylindrical vanadium can of 6 mm diameter and NPD data were collected at temperatures of 2 and 300 K.

**Results and Discussion:**

The elemental compositions were estimated by quantitative analysis of the EDS spectrum which were recorded randomly at different places in the microstructure of CFTO. Since the EDS detector is not very sensitive to the detection of oxygen content, we have excluded the oxygen composition during EDS analysis and results of the elemental composition of atoms are presented in Table S1 of the supplementary material [41]. It is evident that the average atomic weight percent of Cr:Fe:Ti = (24.6 ± 0.5):(25.1 ± 0.6):(50.3 ± 0.7) are in good agreement with the target composition within the standard deviation. Thus, our EDS analysis confirms the chemical homogeneity of CFTO sample.

The phase purity of CFTO was further verified by XRD measurements. As pointed out in the introduction section, Grey et al. [36] first reported the synthesis of CFTO sample and proposed a monoclinic structure in the space group $P2_1/a$ (SG No. 14, Z = 4) with unit cell parameters: a = 7.032 (3) Å, b = 5.000 (2) Å, c = 14.425 (6) Å and α = γ = $90^0$, β = 116.59 (4)$^0$ based on the detailed x-ray diffraction analysis. The proposition of the monoclinic structure of CFTO has close resemblance with the α-$PbO_2$ type structure. Following Grey et al. [36], we have also refined the XRD pattern using the same space group and the results are depicted in Fig. 1(a). It can be seen that the experimentally observed (red filled dots) and calculated (black continuous line) patterns show satisfactory fit for the monoclinic structure in the $P2_1/a$ space group. The blue line represents the difference pattern of the observed and calculated profiles while the vertical ticks indicate the Bragg peak positions. Our careful analysis of the XRD data also revealed a significant ~15 % anti-site disorder between the metal cation $M_1$ and $M_3$ sites. The refined unit cell parameters are: a = 7.0234 (4) Å, b = 4.9932 (3) Å, c = 14.4008 (8) Å and α = γ = $90^0$, β = 116.598 (3)$^0$. These parameters are in good agreement with those reported by the previous workers [36]. The detailed refined structural parameters are given in Table S2 of the supplementary information [41]. A schematic of the crystal structure of CFTO is shown on the lower panel of Fig. 1(b), where the oxygen atoms form a distorted hexagonal close packed lattice with metal cations occupying the octahedral interstices and give rise to a complex arrangement of octahedra connected by the sharing of corners, faces and edges. The direct metal to metal cations and M-O bond distances are summarized in Table S2 of the supplemental information [41]. Evidently, the metal to metal cations distances lies in the range 2.61 - 3.35 Å. On the other hand, the individual M - O bond distances vary between 1.80 - 2.16 Å for $M_2$, $M_4$ and $M_5$ while the disordered sites possess highly distorted octahedra with bond distances for $M_1$ and $M_3$ varying between 1.79 - 2.26 Å and 1.75 - 2.37 Å, respectively.

Having confirmed the good quality of the CFTO sample, we now proceed to investigate the magnetic behaviour. The temperature dependence of the dc magnetic susceptibility ($\chi$) of CFTO measured under zero-field cooled warming (ZFCW) and field cooled warming (FCW) protocols at an applied field of 50 Oe is shown in the main panel of Fig. 2. For the ZFCW protocol, the CFTO sample was cooled from 300 to 2 K in a zero magnetic field and after proper temperature stabilization at 2 K, a 50 Oe field was applied and magnetization was measured as a function of temperature in the warming cycle. For the FCW protocol, the CFTO sample was cooled in the presence of 50 Oe magnetic field from 300 to 2 K and magnetization data was recorded as a function of temperature in the warming cycle. It is evident from the figure that the ZFCW and FCW susceptibility $\chi$ (T) curves overlap with each other and gradually increases with decreasing temperature from 300 K and the two curves bifurcate at a characteristic temperature called the irreversibility temperature ($T_{irr} \sim 23$ K) while ZFCW $\chi$ (T) curve shows a sharp peak at the freezing temperature $T_f \sim 22$ K. Such a bifurcation of the ZFCW and FCW $\chi$ (T) curves below $T_{irr}$ has been reported in several magnetic systems such as spin/cluster glasses [42,43] and superparamagnets (SPM) [44]. For SPM systems, the FCW χ (T) curve increases continuously below the peak temperature $T_f$ while it shows a plateau and then start increasing for glassy systems [44]. In our case, the FCW χ (T) curve shows plateau over a small temperature range and then increases which is typical signature of the spin/cluster glass state (see inset of Fig. 2(a)). To understand the nature of the magnetic interactions in CFTO, we plot inverse of ZFCW susceptibility (1/χ) as a function of temperature at 50 Oe which is shown in the lower panel (b) of Fig. 2. It is evident that above 240 K, the $\chi^{-1}$ increases linearly with temperature and follows Curie-Weiss (C-W) law:

$$\chi (T) = \left(\frac{C}{T-\theta_{CW}}\right) \quad (1)$$

where C is a Curie constant and $\theta_{CW}$ is Curie-Weiss temperature. A least square fit to the observed data in the temperature range 240-300 K yields C = (3.55 ± 0.02) emu K mol$^{-1}$ Oe$^{-1}$ and $\theta_{CW}$ = - (302 ± 0.2) K. The effective magnetic moment (μ$_{eff}$) calculated from the Curie constant C using the relationship (√3Ck$_B$/N$_A$) comes out to be (5.33 ± 0.02) μ$_B$ which is very close to the theoretically estimated value of 5.00 $\mu_B$ using the relationship $\mu_{eff} = \sqrt{0.5(\mu_{Cr^{3+}})^2 + 0.5(\mu_{Fe^{3+}})^2}$ and considering the magnetic moment of $Cr^{3+}$ (3.87 μ$_B$) and $Fe^{3+}$ (5.92 μ$_B$). A small difference between the theoretically and experimentally observed values of effective moment has been attributed to the limited linear region for Curie-Weiss fitting (240-300 K). We believe that the higher temperature χ (T) data well above 300 K may provide better matching of the theoretical and experimental effective moments. The large negative θ$_{CW}$ indicates the strong antiferromagnetic exchange interactions but there is no indication of any long-range ordered antiferromagnetic transition in CFTO. Further, the deviation from the Curie-Weiss behaviour is notable below 240 K which point towards the formation of short-range ordered antiferromagnetic correlations in the CFTO.

In general, the frustration, disorder and randomness in a magnetic system leads to spin-glass state. The frustration parameter [20] can be estimated from the empirical formula f

= $|\Theta_{CW}|/T_c$, where $T_c$ is the ordering transition temperature. For non-frustrated antiferromagnetic system, $\Theta_{CW} \sim T_c$ or $T_N$ and f ≈ 1 but for frustrated AFM system f has value much larger than 1. Generally, f > 5 indicates strong frustration. The value of f for CFTO comes out to be ~ 11 which reveals frustrated exchange interactions in the system. Such frustration has been attributed to the competition between nearest neighbour (NN) and next nearest neighbour (NNN) interactions [20].

To get deeper insights into the origin of the glassy phase in CFTO, we have carried out ac susceptibility ($\chi'$ (T, ω)) measurements as a function of temperature and frequency. Panel (c) of Fig. 2 depicts the temperature dependence of the real part of ac susceptibility at different frequencies with a constant ac drive field of 4 Oe. It is evident that $\chi'$ (T) exhibits a peak across the $T_f$, similar to that observed in dc χ (T) (shown in panel (a) of Fig. 2). Notably, the peak position of $T_f$ shift towards higher temperature side with increasing the frequency of measurement (f). We also note that the magnitude of $\chi'$ (T) decreases with increasing frequency. Such a frequency dependent peak shift of $T_f$ corresponding to $\chi'$ (T, ω) can be due to spin /cluster freezing of superparamagnetic (SPM) blocking [42–44]. In order to distinguish the two phenomena, we analyse the empirical Mydosh parameter (Φ) given as [42]:

$$\Phi = \Delta T_f /(T_f \Delta ln(\omega)), \qquad (2)$$

where $\Delta T_f$ represent the shift in the peak temperatures corresponding to highest and lowest frequencies. The calculated value of Φ for CFTO comes out to be ~ 0.003 which is orders of magnitude lower than that reported for the SPM systems (0.10 - 0.30) and belongs to the expected range of spin/cluster glass systems (0.005 - 0.09) [42–44]. Thus our Mydosh parameter analysis suggests the glassy freezing in CFTO.

The M-H hysteresis curves recorded under the zero-field cooled condition at some selected temperatures are depicted in Fig.3. It is evident that the M-H curve at 300 is linear in nature, which confirms the paramagnetic behaviour. We notice that at 2 K the M-H hysteresis loop does not saturate at 7 T field revealing the antiferromagnetic nature of the CFTO. It can be clearly seen that below the peak temperature (i.e. $T < T_f$) a significant non-linearity and slim hysteresis loop with small remnant magnetization (for example at 2 K, $H_c \sim 1060$ Oe, and $M_r \sim 0.2$ emu/g) are observed which is expected in glassy systems [45–47].

To gain more insights on the magnetic ground state of CFTO, we performed specific heat measurements as function of temperature under zero applied magnetic field. The result is shown in the main panel of Fig. 4. Evidently, $C_p$ vs T plot does not exhibit any anomaly (typically, a λ like feature is observed for long-range ordered transition [48]) down to 2 K. This confirms that long-range magnetic ordering and/or a structural phase transition is absent in CFTO. There is no non-magnetic analogue for CFTO which have the same crystal structure and whose specific heat can be used to subtract the lattice contribution ($C_{lattice}$) for obtaining the magnetic contribution ($C_m$). The $C_{lattice}$ contribution for CFTO was thus estimated by fitting the $C_p$ versus T plot in the temperature range 75-200 K with combined Debye and Einstein function of the specific heat [48] and then extrapolated the fitted curve down to lowest temperature 2 K. The $C_{lattice}(T)$ can be expressed as:

$$C_{lattice}(T) = (1-\alpha) C_{Debye} + \alpha C_{Einstein} \qquad (3)$$

$$C_{Debye} = 9nR \left(\frac{T}{\Theta_D}\right)^3 \int_0^{\Theta_D/T} \frac{x^4 e^x}{(e^x-1)^2} dx, \quad x = \frac{\hbar\omega}{k_B T} \qquad (4)$$

$$C_{Einstein} = 3nR \left(\frac{\Theta_E}{T}\right)^2 \frac{e^{\frac{\Theta_E}{T}}}{(e^{\frac{\Theta_E}{T}}-1)^2} \qquad (5)$$

where $\alpha$ is the fraction of Einstein contribution to the total specific heat, $\Theta_D$ and $\Theta_E$ are Debye and Einstein temperatures, n is the number of atoms per formula unit, $R$ is a universal gas constant, $h$ and $k_B$ are Planck's and Boltzmann constant, ω is the phonon frequency. The least-square fitting of the observed total $C_p(T)$ data in the T-range 75-200 K yields: $\alpha = 0.552$, $\Theta_D \approx$ 390 K and $\Theta_E \approx 690$ K. The $C_{lattice}$ is shown as red continuous line in the Fig. 4. The $C_m$ (T) contribution of CFTO was estimated by subtracting the $C_{lattice}$ (T) from the total $C_p$ (T). The temperature variation of $C_m$ (T) is shown in the lower inset of Fig. 4. It is evident that $C_m$ increases with decreasing temperature and displays a broad peak ~ 35 K which is ~ 1.5 times of $T_f$ consistent with the well-known dilute spin-glasses [42,43]. Interestingly, $C_m$ shows a linear temperature dependence below $T_f$ (see lower inset of Fig. 4). Such a linear temperature dependence of $C_m$ has been reported in spin glasses and explained theoretically using a two level tunnelling model [49].

The magnetic entropy ($S_m$) release can be determined by integrating the $C_m/T$ over a 2 to 200 K range. The variation of $S_m$ as a function of temperature is shown in the upper inset of Fig. 4. The total expected magnetic entropy ($S_m$) of the CFTO system [0.5*R {ln(2S($Cr^{3+}$; S = 3/2) +1)} + 0.5*R {ln(2S ($Fe^{3+}$; S = 5/2) +1}] is ~13.21 J/mol-K. The maximum observed value of $S_m$ for CFTO is ~5.8 J/mol-K, which is approximately 44% of the total expected $S_m$. The value of entropy at $T_f$ is found to be only 2.06 J/mol-K (see upper inset of Fig. 4) which is 15 % of the total expected magnetic entropy. This suggests that the large amount of magnetic entropy is contained in the short-range correlations which are developed much above the freezing temperature. Thus, our specific heat analysis also confirms the glassy ground state in CFTO consistent with the dc and ac susceptibility studies.

Neutron scattering measurements has proven to be a powerful tool for studying the nature of magnetic ordering and magnetic correlations in both ordered and disordered systems [50]. Keeping this in mind, we have performed neutron powder diffraction (NPD) measurements using a powder sample of CFTO. Fig. 5 depicts the NPD pattern of CFTO at selected temperatures 1.7 K and 300 K over a 2θ range of 2 to 85º. It can be seen from the figure that no magnetic Bragg peaks are observed in the NPD patterns down to 1.7 K, however, a broad diffuse feature is seen around 2θ ~32º. This broad feature arises due to magnetic diffuse scattering, as a result of the development of the short-range ordered antiferromagnetic (AFM) correlated clusters. It is also evident that the broad diffuse peak is present even at 300 K and gets more pronounced at low temperatures i.e., at 1.7 K (see inset (a) of Fig. 5 on a zoomed scale). This indicates that the short-range ordered AFM correlations starts to develop close to room temperature itself. A similar broad diffuse peak has been reported in spin-chain compounds [51], honeycomb layered systems [52], spinels [53], complex perovskite [54,55] and in some geometrically frustrated pyrochlore systems [56].

In order to better resolve the magnetic diffuse scattering, we have subtracted the background of NPD data taken at 300 K to 2 K data. After subtraction, the diffuse magnetic peak at 1.7 K was least square fitted with the Gaussian function and is shown in the inset (b) of Fig. 5. This magnetic diffuse peak allows us to get a rough estimate of the magnetic correlation length ($\xi$) using the Scherrer formula:

$$\xi = 0.9\lambda/\beta Cos\theta \qquad (6)$$

where $\beta$ is the full-width half maxima (FWHM) and $\lambda$ is the wavelength used for the neutron scattering measurements (2.458 Å). To determine the intrinsic FWHM, we have corrected the observed FWHM with the instrumental broadening. The correlation length $\xi$ so-obtained is ~ 20 Å at 1.7 K. The short $\xi$ reveals that the highly restricted nature of AFM interactions present in the system. We believe that the broad peak corresponds to the nearest neighbour spin correlations and the large peak width indicates that the spin correlations exist over a short-range scale. Our neutron scattering studies thus provide microscopic evidence for the presence of short-range ordered AFM correlated spin clusters in CFTO with average size of ~20 Å which seem to be involved in the glassy dynamics.

To further probe the glassy ground state and roughly estimate the spin-glass correlation length, we perform systematic iso-thermal remanent magnetization and aging experiments. The measurement protocol consists of cooling the sample in a magnetic field through the glassy freezing temperature ($T_f$) to a desired measuring temperature $T$ (< $T_f$), waiting a time $t_w$, then switched-off the field to zero and recording the decay of magnetization ($M_{IRM}$) as a function of time. The results for a fixed $t_w$ and different field $H$ are shown in Fig. 6. It is evident that the $M_{IRM}$ curve decays very slowly with increasing time. Such a very slow decay arises due to the presence of the metastable states which are separated by distinct energy barriers. After the release of the applied magnetic field, these metastable states respond to the different relaxation times. The relaxation rate $S(t)$ also referred as the 'response function' can be expressed as [57]:

$$S(t) = d\left[\frac{-M_{IRM}(t, t_w)}{H}\right]/d(\ln t) \qquad (7)$$

where, $M_{IRM}$ is the isothermal remanent magnetization at time t for a waiting time $t_w$ after removing the field to zero. Fig. 6(b) show the time evolution of the $S(t)$ at different applied magnetic fields. The following features are clearly discernible from the figure: (1) $S(t)$ exhibits a peak at the effective waiting time $t_w^{eff}$ (less than the actual $t_w$), (2) the peak position of $S(t)$ shift towards lower time scale with increasing field. Experimentally, the peak in the $S(t)$ corresponding to effective wait time $t_w^{eff}$ has been reported in the glassy systems like $Eu_{0.5}Ba_{0.5}MnO_3$ [58], Cu:Mn 6 at.% [57] and the insulating thiospinel $CdCr_{1.7}In_{0.3}S_4$ [57].

The magnetic field dependence of the peak of $S(t)$ has been used to roughly estimate the spin-glass correlation length [57]. The relaxation rate $S(t)$ is directly related to the typical value of the free-energy barriers. When the magnetic field is increased, the effective barrier heights are reduced by Zeeman energy term $E_z$. The effective wait time (at which $S(t)$ peaks) $t_w^{eff}$ is smaller than the actual wait time $t_w$ and is related by $\ln(t_w^{eff}/t_w) = - E_z/k_BT$ [59]. Further, the Zeeman energy $E_z \approx \sqrt{N}H\,m\mu_B$, where N represents the number of correlated spins which are locked in the barrier hopping process, H is the applied field, m is the effective magnetic moment and $\mu_B$ is the Bohr magneton. The value of N depends on both $t_w$ and T and is roughly related to the correlation length by $\xi(t_w, T) \approx [N(t_w, T)]^{1/3}$ [57]. The estimated value of N for CFTO is of the order of ~ 4300, which is consistent with that reported in $Fe_{0.5}Mn_{0.5}TiO_3$ system but an order of magnitude smaller than that reported in the systems like $Eu_{0.5}Ba_{0.5}MnO_3$ [58], Cu:Mn 6 at.% [57], Ag:Mn 2.7at.% [59] and insulating thiospinel $CdCr_{1.7}In_{0.3}S_4$ [57]. The calculated effective correlation length $\xi$ for CFTO comes out to be of the order of ~16Å which is very close to ~20 Å obtained by our neutron diffraction studies. A similar order of $\xi$ has been reported in other spin-glass systems $La_{1-x}Sr_xCoO_3$ ($\xi \approx 25$ Å) [60], $Eu_{0.5}Ba_{0.5}MnO_3$ ($\xi \approx 35$ Å) [58], $MnSb_2Se_4$ ($\xi \approx 20$ Å) [61], $Fe_2O_3$ ($\xi \approx 30$ Å) [62], $CoCr_2O_4$ ($\xi \approx 20$ Å) [63] and $CaNi_3P_4O_{14}$ ($\xi \approx 4.8$ Å) [51]. Thus, our detailed analysis of the isothermal remanent magnetization further support the existence of short-range ordered spin clusters in CFTO.

## Conclusions:

In summary, we have successfully synthesized the $CrFeTi_2O_7$ (CFTO) system and characterized it using both macroscopic and microscopic probes. A careful analysis of the X-ray diffraction data confirms that CFTO crystallizes in the monoclinic crystal structure in the $P2_1/a$ space group. Despite very strong antiferromagnetic interactions between the magnetic Cr and Fe ions, as indicated by large negative Curie-Weiss constant $\theta_{CW} = -(302 \pm 0.2)$ K, no signature of long range ordering is observed down to 2 K. However, the temperature dependent zero-field cooled warming (ZFCW) and field cooled warming (FCW) dc susceptibility measurements indicate the glassy freezing at $T_f \sim 22$ K. This is further supported by ac susceptibility data which shows frequency dispersion across, isothermal magnetization and specific heat measurements. Neutron powder diffraction (NPD) analysis not only reveals the absence of any long range ordering in CFTO but also provides microscopic evidence for the presence of short-range ordered antiferromagnetic correlated spin clusters with average size of ~20 Å. This is further reconfirmed from iso-thermal remanent magnetization measurements.

## Acknowledgments:

Arun Kumar acknowledges SERB-DST, Government of India and I-HUB Quantum Technology Foundation Pune for providing financial support through a Post-Doctoral Fellowship (PDF/2020/002116). SN acknowledges support from an Air Force Research Laboratory Grant (FA2386-21-1-4051).

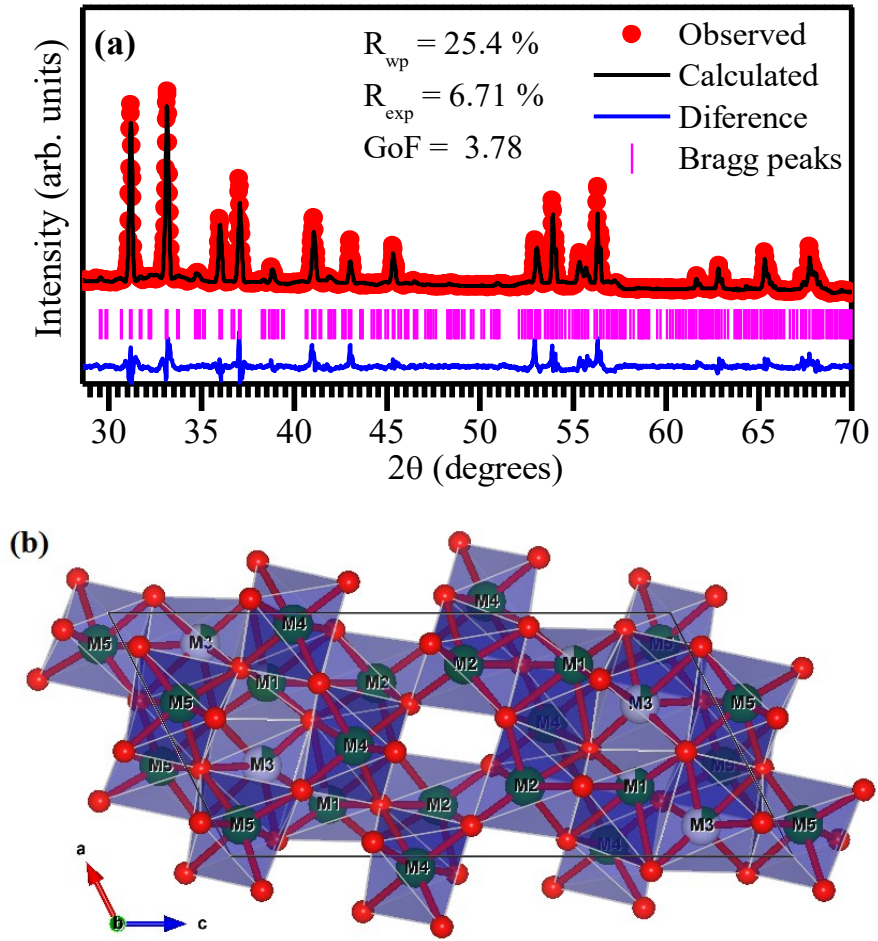

**FIG. 1:** (a) Observed (red filled dots), calculated (black solid line) and difference (blue solid line) patterns obtained after the refinement of x-ray diffraction data of *CrFeTi$_2$O$_7$* at room temperature using *P2$_1$/a* space group. Vertical tick marks above the difference line represent the Bragg peak positions. (b) A schematic of the crystal structure of *CrFeTi$_2$O$_7$* as viewed down [010] direction. The M$_3$ and M$_1$ represent the disorder sites.

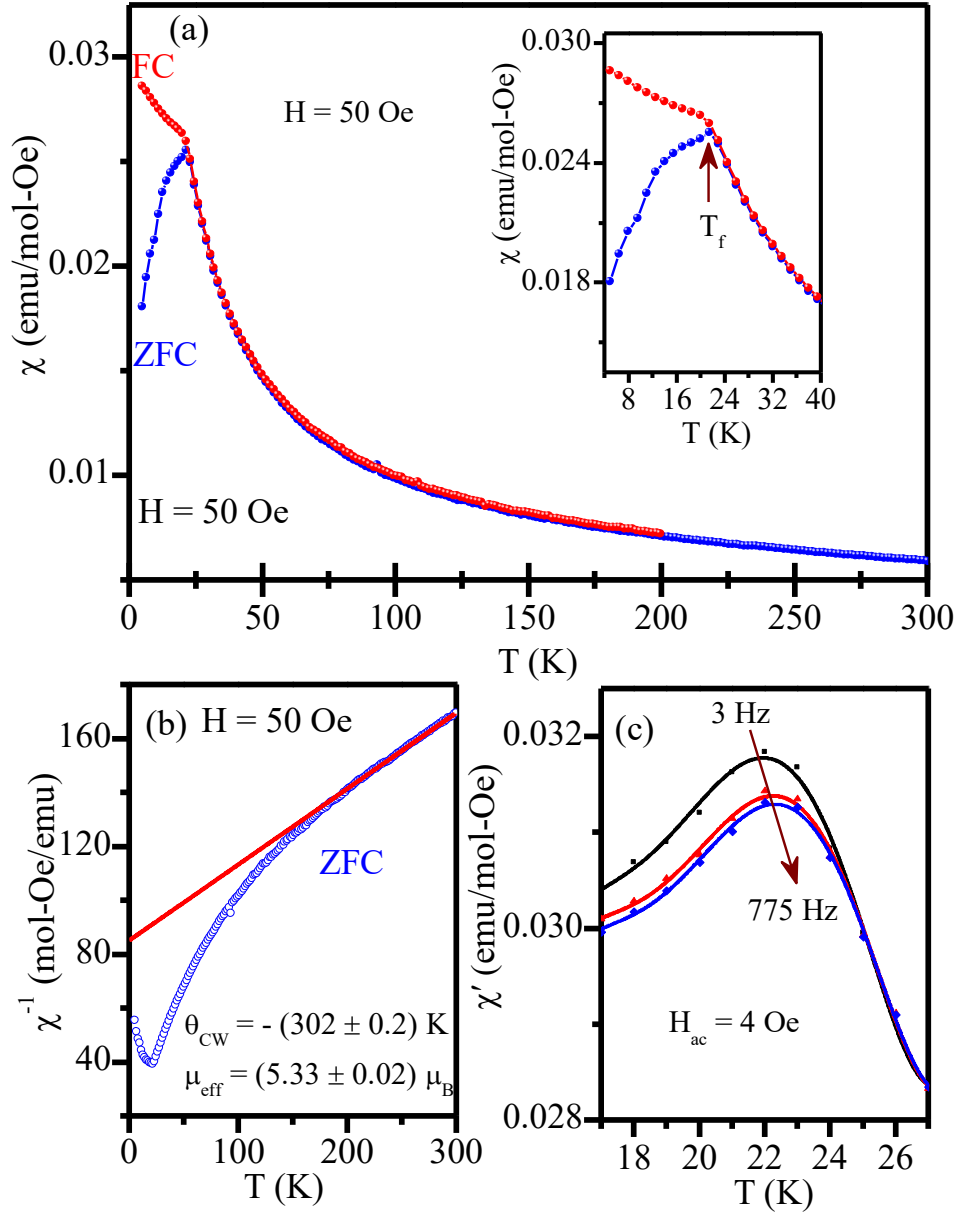

**FIG. 2:** Panel (a) Temperature-dependent zero-field cooled warming (ZFCW) and field cooled warming (FCW) dc susceptibility of *CrFeTi$_2$O$_7$* measured at 50 Oe field. Insets of panel (a) show the magnified scale of ZFCW and FCW dc susceptibility across the freezing temperature. Panel (b) depicts the inverse of ZFCW dc susceptibility ($\chi^{-1}$) as a function of temperature for 50 Oe applied field and solid line through data points represents the Curie-Weiss fitting in the 240-300 K range. Panel (c) shows the temperature variation of the real part of ac magnetic susceptibility (($\chi'(\omega, T)$) of *CrFeTi$_2$O$_7$* measured at different frequencies 3 Hz, 132 Hz and 775 Hz with constant ac excitation field of 4 Oe.

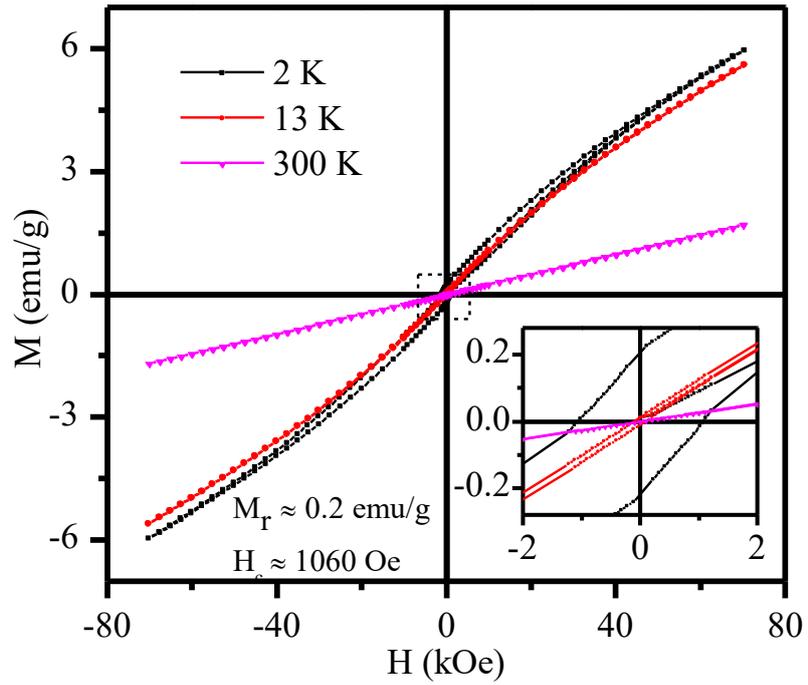

**FIG.3**: Magnetization versus magnetic field curves (M-H) measured at some selected temperatures after cooling under zero field cooled conditions. Inset depicts the magnified view of the M-H curve at low field.

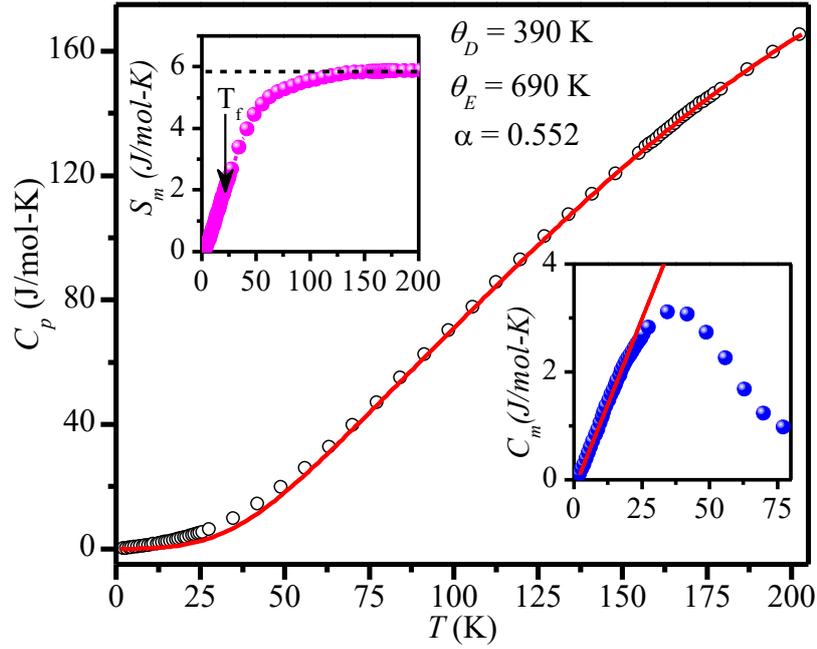

**FIG 4:** The main panel depicts the temperature dependence of total specific heat ($C_p$: open circle) of *CrFeTi$_2$O$_7$* at zero magnetic field along with lattice specific heat ($C_{lattice}$: red continuous line) obtained from the fitting of Debye and Einstein models as describe in the text. The lower inset shows the magnetic specific heat ($C_m$: blue filled circle) obtained by subtracting the lattice contribution from the total specific heat. Solid line through data points represent the linear behaviour of $C_m$ below $T_f$. The upper inset depicts the variation of magnetic entropy $S_m$ as a function of temperature.

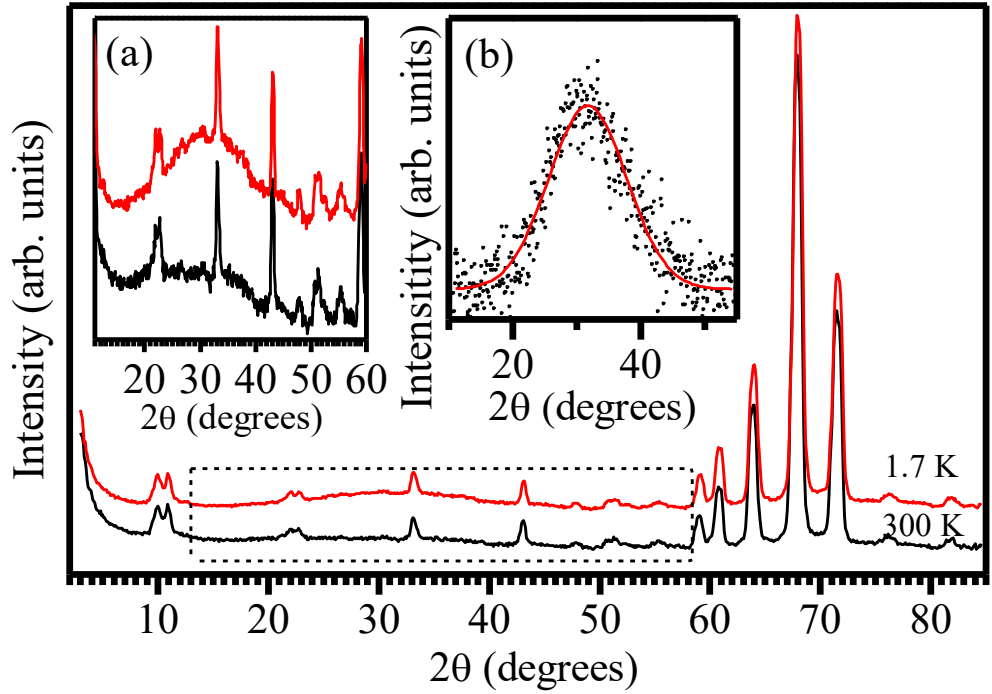

**FIG. 5:** Neutron powder diffraction patterns of *CrFeTi$_2$O$_7$* at selected temperatures 300 K and 2 K. The patterns are shifted vertically for the purpose of presentation. Inset (a) depicts the enlarged scale of broad diffuse magnetic scattering peak corresponding to short-range antiferromagnetic correlations. Inset (b) shows the diffuse magnetic peak at 1.7 K after the background subtracted from the 300 K NPD data and solid line is the fit for the Gaussian function.

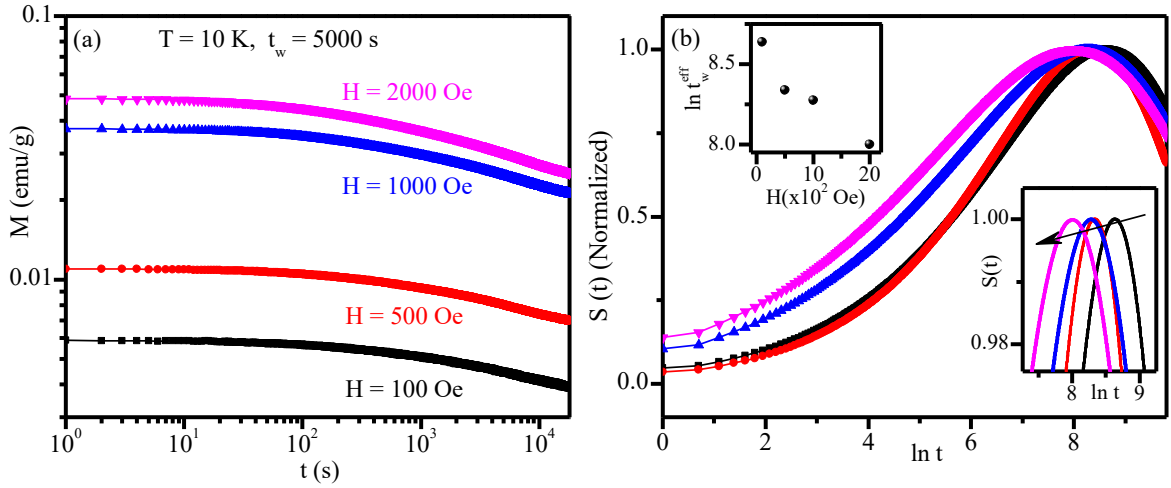

**FIG. 6:** Panel (a) show the time evolution of isothermal remanent magnetization $M_{IRM}$ at 10 K for a fixed waiting time $t_w$ of 5000 s with various applied magnetic fields. Panel (b) show the $S(t)$ versus $t$ plot at different magnetic fields. The lower inset of (b) depicts the observed peak in the $S(t)$ on a magnified scale. The upper inset of (b) shows the $\ln t_w^{\text{eff}}$ versus $H$ plot.